\documentclass[letterpaper,twocolumn,english,showpacs,preprintnumbers,amsmath,amssymb,superscriptaddress,nofootinbib,pra]{revtex4}
\usepackage[T1]{fontenc}
\usepackage[latin9]{inputenc}
\usepackage{amsmath}
\usepackage{amssymb}
\usepackage{graphicx}
\usepackage{setspace}
\usepackage{esint}

\makeatletter

\pdfpageheight\paperheight
\pdfpagewidth\paperwidth

\@ifundefined{textcolor}{}
{%
 \definecolor{BLACK}{gray}{0}
 \definecolor{WHITE}{gray}{1}
 \definecolor{RED}{rgb}{1,0,0}
 \definecolor{GREEN}{rgb}{0,1,0}
 \definecolor{BLUE}{rgb}{0,0,1}
 \definecolor{CYAN}{cmyk}{1,0,0,0}
 \definecolor{MAGENTA}{cmyk}{0,1,0,0}
 \definecolor{YELLOW}{cmyk}{0,0,1,0}
}

\renewcommand{\vec}[1]{\mathbf{#1}}
\renewcommand{\Re}{\operatorname{Re}}
\renewcommand{\Im}{\operatorname{Im}}

\usepackage{babel}

\makeatother

\usepackage{babel}
\begin{document}

\title{Optical ``Bernoulli'' forces}

\author{Ramis Movassagh }

\email{ramis.mov@gmail.com}

\affiliation{Department of Mathematics, Northeastern University, Boston MA, 02115}

\author{Steven G. Johnson}

\affiliation{Department of Mathematics, Massachusetts Institute of Technology,
Cambridge MA, 02139}

\date{\today}
\begin{abstract}
\begin{singlespace}
By Bernoulli's law, an increase in the relative speed of a fluid around
a body is accompanies by a decrease in the pressure. Therefore, a
rotating body in a fluid stream experiences a force perpendicular
to the motion of the fluid because of the unequal relative speed of
the fluid across its surface. It is well known that light has a constant
speed irrespective of the relative motion. Does a rotating body immersed
in a stream of photons experience a Bernoulli-like force? We show
that, indeed, a rotating dielectric cylinder experiences such a lateral
force from an electromagnetic wave. In fact, the sign of the lateral
force is the same as that of the fluid-mechanical analogue as long
as the electric susceptibility is positive ($\epsilon>\epsilon_{0}$),
but for negative-susceptibility materials (e.g. metals) we show that
the lateral force is in the opposite direction. Because these results
are derived from a classical electromagnetic scattering problem, Mie-resonance
enhancements that occur in other scattering phenomena also enhance
the lateral force.\end{singlespace}

\end{abstract}
\maketitle
\begin{singlespace}
\textit{\label{sec:Bernoulli_intro}Photonic Bernoulli's Law?} When
considering a rotating body in a fluid stream such as air, the body
experiences a pressure gradient caused by the difference of the relative
velocity of its motion to that of the fluid at various points on its
boundary. For example, an idealized tornado such as a spinning cylinder
moves perpendicular to the streamlines of the fluid. The direction
of motion is along the direction connecting the center of the cylinder
to the point of maximum relative velocity. 
\end{singlespace}

In a famous experiment, Michelson and Morley \cite{Moller1957} showed
that even if the earth were immersed in a fluid in motion, the speed
of light would be constant relative to perpendicular directions. Later,
the special theory of relativity established the constancy of the
speed of light regardless of observer's relative motion to the light
source. Here, we ask to what extent can a stream of photons resembles
a stream of massive fluids? In particular, if one considers a stream
of photons (classically described by Maxwell's equations) as a fluid
in motion and places a rotating dielectric body in it, one might naively
expect that no Bernoulli-type force would be experienced by the body
since the relative speed of light is the same on both sides. Here
we show that such a force \textit{is} experienced by the rotating
body, though the cause is the asymmetry of the scattered field \cite{Tai1964}
from the dielectric, by which a net force is imparted to the rotating
body. 
\begin{figure}
\begin{centering}
\includegraphics[scale=0.18]{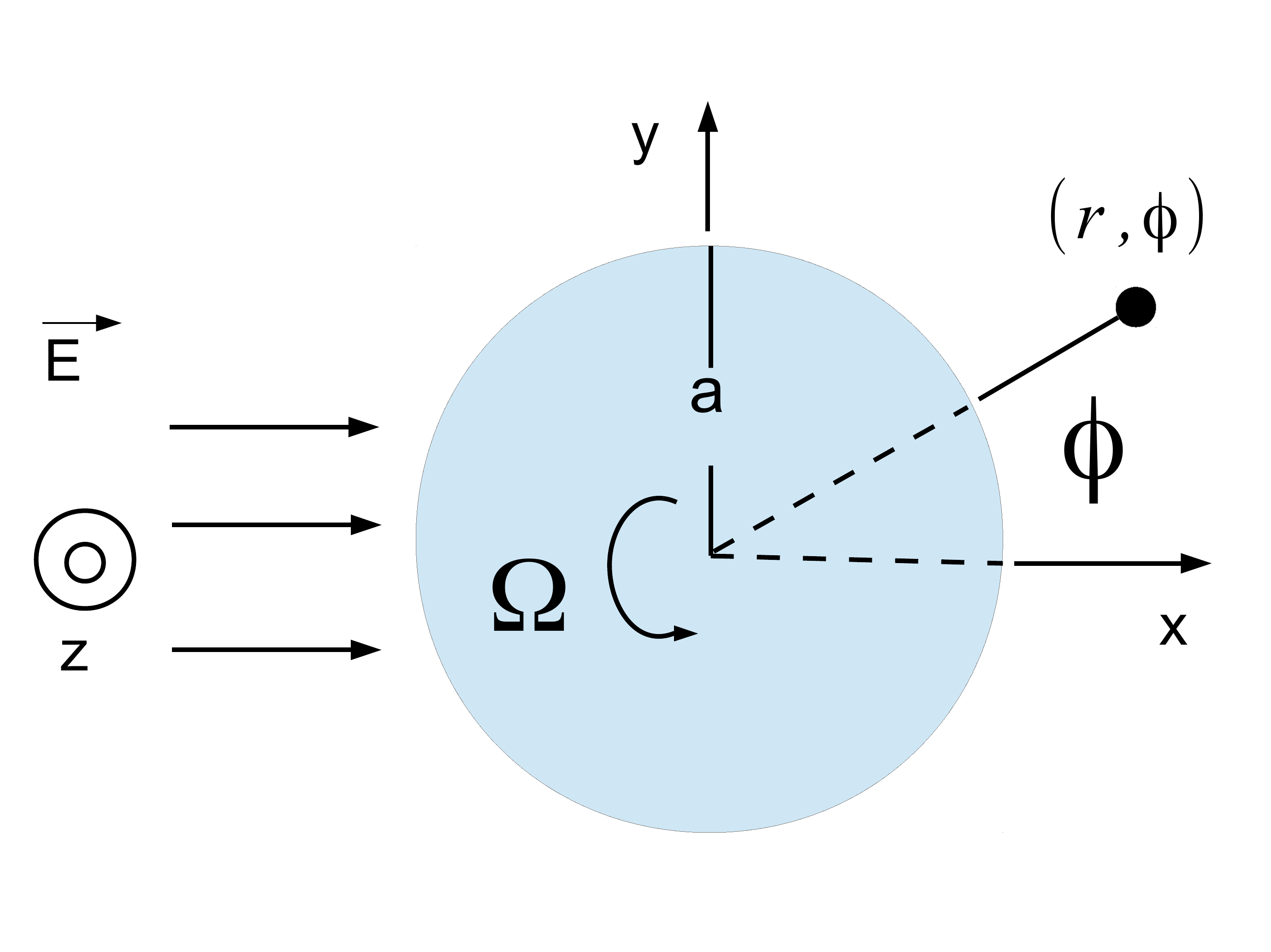} 
\par\end{centering}

\caption{\label{fig:CylindricalGeometry}(Color online) Light scattering from
a rotating dielectric cylinder.}
\end{figure}

\begin{singlespace}
\textit{\label{sec:ScatteringField} Cylindrically Rotating Dielectric.--}
The exact electromagnetic constitutive equations in a medium moving
at velocity $\vec{v}$, discovered by Minkowski \cite{Minkowski1907},
are

\begin{eqnarray}
\vec{D}+\vec{v}\times\vec{H}/c & = & \epsilon\left(\vec{E}+\vec{v}\times\vec{B}/c\right)\label{eq:FundamentalEquationsDielectricsExact}\\
\vec{B}+\vec{E}\times\vec{v}/c & = & \mu\left(\vec{H}+\vec{D}\times\vec{v}/c\right),\label{eq:FundamentalEquationsDielectricsExact2}
\end{eqnarray}
 where $\vec{E}$, $\vec{D}$, $\vec{B}$ and $\vec{H}$ are the usual
electromagnetic fields, $c$ is the speed of light in vacuum and $\epsilon$
is the electric permittivity in the rest frame and $\mu$ is the magnetic
permeability in the rest frame. These equations presuppose uniform
motion of the dielectric, where special relativity is sufficient.
For \textit{accelerated} dielectrics, the equations become more complicated;
however, for rotating bodies with axial symmetry, the body in motion
has the same shape as the one in the rest frame and it has been shown
that the same equations would apply~\cite{Sommerfeld1952,Ridgely1996,Tai1964}.
This assertion has been successfully used in applications~\cite{Tai1964}
and was later proved rigorously by Ridgely~\cite{Ridgely1996}, who
showed that the general relativistic treatment for \textit{uniformly
rotating} dielectrics with axial-symmetry, to first order in $v/c$,
gives Minkowski's results (Eqs.~\ref{eq:FundamentalEquationsDielectricsExact}
and \ref{eq:FundamentalEquationsDielectricsExact2}). 
\end{singlespace}

In the limit where $v/c$ is small, Tai considered the scattered field
of a plane wave incident upon a uniformly rotating dielectric cylinder
with angular speed $\Omega$ \cite{Tai1964}. We begin by reviewing
Tai's derivation of the scattered field and then we use these fields
to compute the force. As depicted in Fig.~\ref{fig:CylindricalGeometry},
the velocity of the rotating body is $\vec{v}=\Omega r\vec{\hat{{\phi}}}$
at a radius~$r$, the radius of the cylinder is denoted by $a$,
and the $\vec{E}$ field of the incident wave is assumed to be polarized
in the direction of the axis of the cylinder (which we take to be
$\mathbf{\hat{z}}$).

Derivations of key equations are provided in the appendix.

We solve the scattering problem by standard technique of expanding
the field in each region in basis of Bessel functions $J_{n}$ and
then matching boundary conditions at the interface. In this basis,
an incident $z$-polarized plane wave propagating in the $+x$ direction
with amplitude $E_{0}$ (see Fig.~\ref{fig:CylindricalGeometry})
is given in polar $(r,\phi)$ coordinates by

\begin{singlespace}
\begin{equation}
E_{i}=E_{0}\mbox{exp}\left\{ ik_{0}r\cos\phi\right\} =E_{0}\sum_{n=-\infty}^{+\infty}i^{n}J_{n}\left(k_{0}r\right)\mbox{exp}\left(in\phi\right),\label{eq:IncidentE_Tai}
\end{equation}
 where $E_{0}$ is the amplitude, $k_{0}=\omega/c$ is the wave number
in vacuum, $\omega$ is the frequency in the time-harmonic oscillating
field $e^{-i\omega t}$. The scattered and ``transmitted'' (interior)
fields, respectively, can be written (using the Hankel function $H_{n}^{(1)}=J_{n}+iY_{n}$)
with to-be-determined coefficients $\alpha_{n}$ and $\beta_{n}$:

\begin{eqnarray}
E_{s} & = & E_{0}\sum_{n=-\infty}^{+\infty}\alpha_{n}i^{n}H_{n}^{\left(1\right)}\left(k_{0}r\right)\mbox{exp}\left(in\phi\right).\label{eq:ScatteredE_Tai}\\
E_{t} & = & E_{0}\sum_{n=-\infty}^{+\infty}\beta_{n}i^{n}J_{n}\left(\gamma_{n}r\right)\mbox{exp}\left(in\phi\right),\label{eq:E_t_RAW}
\end{eqnarray}
 where $\gamma_{n}$ is defined by 
\end{singlespace}

\begin{eqnarray}
K & \equiv & \mu_{0}\left(\epsilon-\epsilon_{0}\right)\Omega\label{eq:K}\\
m & \equiv & 1-\frac{\epsilon_{0}}{\epsilon}\\
\gamma_{n}^{2} & = & k^{2}-2n\omega K=k^{2}\left(1-\frac{2nm\Omega}{\omega}\right).\label{eq:gamma_n}
\end{eqnarray}
 The total field is therefore $\mathbf{E}=E_{z}\hat{\mathbf{z}}=\left(E_{i}+E_{s}\right)\hat{\mathbf{z}}$
and the magnetic field in the vacuum regions is given by $\vec{H}=\frac{1}{i\omega\mu_{0}}\nabla\times\vec{E}$.

The unknown coefficients $\beta_{n}$ and $\alpha_{n}$ are found
by requiring continuity of $E_{z}$ and $H_{\phi}$ at $r=a$, yielding

\begin{singlespace}
\begin{eqnarray}
J_{n}\left(k_{0}a\right)+\alpha_{n}H_{n}^{\left(1\right)}\left(k_{0}a\right) & = & \beta_{n}J_{n}\left(\gamma_{n}a\right)\label{eq:BC_alpha_beta}\\
k_{0}\left[J_{n}'\left(k_{0}a\right)+\alpha_{n}H_{n}^{\left(1\right)}\left(k_{0}a\right)'\right] & = & \beta_{n}\gamma_{n}J_{n}'\left(\gamma_{n}a\right),\label{eq:BC_alpha_beta2}
\end{eqnarray}
 where the prime on $J$ and $H^{\left(1\right)}$ denotes derivative
with respect to the entire argument. Solving for $\alpha_{n}$ gives 
\end{singlespace}

\begin{widetext}

\begin{equation}
\alpha_{n}=-\frac{J_{n}\left(\rho_{0}\right)\left[J_{n-1}\left(\rho_{n}\right)-J_{n+1}\left(\rho_{n}\right)\right]-\frac{k_{0}}{\gamma_{n}}J_{n}\left(\rho_{n}\right)\left[J_{n-1}\left(\rho_{0}\right)-J_{n+1}\left(\rho_{0}\right)\right]}{H_{n}^{\left(1\right)}\left(\rho_{0}\right)\left[J_{n-1}\left(\rho_{n}\right)-J_{n+1}\left(\rho_{n}\right)\right]-\frac{k_{0}}{\gamma_{n}}J_{n}\left(\rho_{n}\right)\left[H_{n-1}^{\left(1\right)}\left(\rho_{0}\right)-H_{n+1}^{\left(1\right)}\left(\rho_{0}\right)\right]}.\label{eq:alphan_simplified}
\end{equation}

\end{widetext}where $\rho_{0}=k_{0}a$ and $\rho_{n}=\gamma_{n}a$.
For $\Omega\ne0$, the rotation breaks the $y=0$ mirror symmetry
leading to asymmetrical scattering $\alpha_{n}\ne\alpha_{-n}$ as
shown by Tai \cite{Tai1964}. If $\Omega=0$, then $\alpha_{n}=\alpha_{-n}$
and Eq. \ref{eq:ScatteredE_Tai} reduces to symmetrical scattering.

\begin{singlespace}
\label{sec:Force}\textit{Force imparted to the Rotating Cylinder.--}
The asymmetry in the momentum transport by the scattered field should
manifest itself as a lateral force on the dielectric. This force can
be computed by integrating the Maxwell stress tensor over a closed
surface around the object. Because we only evaluate the stress tensor
in vacuum, we avoid the well-known difficulties that arise in defining
the stress tensor inside the material~\cite{LandauVol8(2ndEdition)},
nor does the rotation affect the vacuum stress tensor. The stress
tensor in SI units is 
\end{singlespace}

\begin{eqnarray}
\overleftrightarrow{\vec{\sigma}} & = & \epsilon_{0}\vec{E}\otimes\vec{E}+\mu_{0}\vec{H}\otimes\vec{H}\label{eq:StressTensorSI}\\
 & - & \frac{1}{2}\left(\epsilon_{0}E^{2}+\mu_{0}H^{2}\right)\left(\vec{\hat{x}}\otimes\vec{\hat{x}}+\vec{\hat{y}}\otimes\vec{\hat{y}}+\vec{\hat{z}}\otimes\vec{\hat{z}}\right)\nonumber 
\end{eqnarray}
 where the hatted quantities are unit vectors. To calculate the force
on the cylinder in any direction $\hat{\mathbf{n_{0}}}$ on the plane
at a fixed radius $r_{0}$, we evaluate

\begin{singlespace}
\begin{equation}
F_{\vec{\hat{n_{0}}}}=\frac{\omega}{2\pi}\int_{0}^{\frac{2\pi}{\omega}}dt\oint r\, dr\, d\phi\,\delta\left(r-r_{0}\right)\left\{ \hat{\mathbf{n_{0}}}\cdot\overleftrightarrow{\vec{\sigma}}\cdot\hat{\mathbf{r}}\right\} ,\label{eq:Force}
\end{equation}
 where $\hat{\mathbf{r}}=\cos\phi\vec{\hat{x}}+\sin\phi\vec{\hat{y}}$
and the time average is taken over a full period to obtain a real
force. For our polarization, $E_{x}=E_{y}=H_{z}=0$. The force in
$\hat{y}$ direction is

\begin{eqnarray}
F_{\vec{\hat{y}}} & = & r_{0}\oint d\phi\left\{ -\frac{\epsilon_{0}\left|E_{z}\right|^{2}+\mu_{0}\left|\vec{H}\right|^{2}}{4}\sin\phi\right.\label{eq:ForceY}\\
 & + & \left.\frac{\mu_{0}}{2}\mbox{Re}\left(H_{x}^{*}H_{y}\right)\cos\phi+\frac{\mu_{0}\left|H_{y}\right|^{2}}{2}\sin\phi\right\} .\nonumber 
\end{eqnarray}
 Using orthogonality conditions on $e^{in\phi}$, Eq.~\ref{eq:ForceY}
can be integrated analytically. In the figures below we use the dimensionless
force $\frac{F_{\vec{\hat{n_{0}}}}c}{P}$, where $P=2a\left|E_{0}\right|^{2}\sqrt{\frac{\epsilon_{0}}{\mu_{0}}}$
is the incident power on the scatterer's geometric cross section.
This is a convenient normalization because, for light incident on
a perfectly absorbing flat surface the force is exactly $\frac{P}{c}$,
so this normalization gives a measure of the lateral force relative
to the incident photon pressure. Furthermore, we use the dimensionless
angular frequency $\frac{a\Omega}{c}$, which is the ratio of speed
of the cylinder boundary to that of light. 
\end{singlespace}

\begin{figure}
\centering{}\includegraphics[scale=0.3]{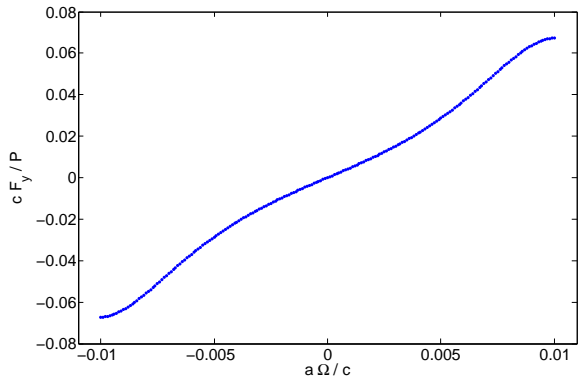}\caption{\label{fig:Force-vs.-rotational}(Color online) Normalized force vs.
rotational frequency for $\lambda_{0}=0.01$ and $\epsilon/\epsilon_{0}=10$
where $P=2a\left|E_{0}\right|^{2}\sqrt{\frac{\epsilon_{0}}{\mu_{0}}}$.}
\end{figure}

Fig. \ref{fig:Force-vs.-rotational} plots the force vs. angular frequency
for $\epsilon/\epsilon_{0}=10$. As discussed below, whenever $\epsilon>\epsilon_{0}$,
corresponding to a positive electric susceptibility $\chi^{(\mbox{e})}=\epsilon/\epsilon_{0}-1$,
there is a force of sign analogous to Bernoulli's law, where $\Omega>0$
(counterclockwise rotation) gives a force in the positive $y$ direction.
Clearly, $\Omega<0$ gives the same magnitude of the force in the
opposite direction as expected from the symmetry of the problem and
in accordance with the fluid-mechanical analogy.

\begin{figure}
\centering{}\includegraphics[scale=0.3]{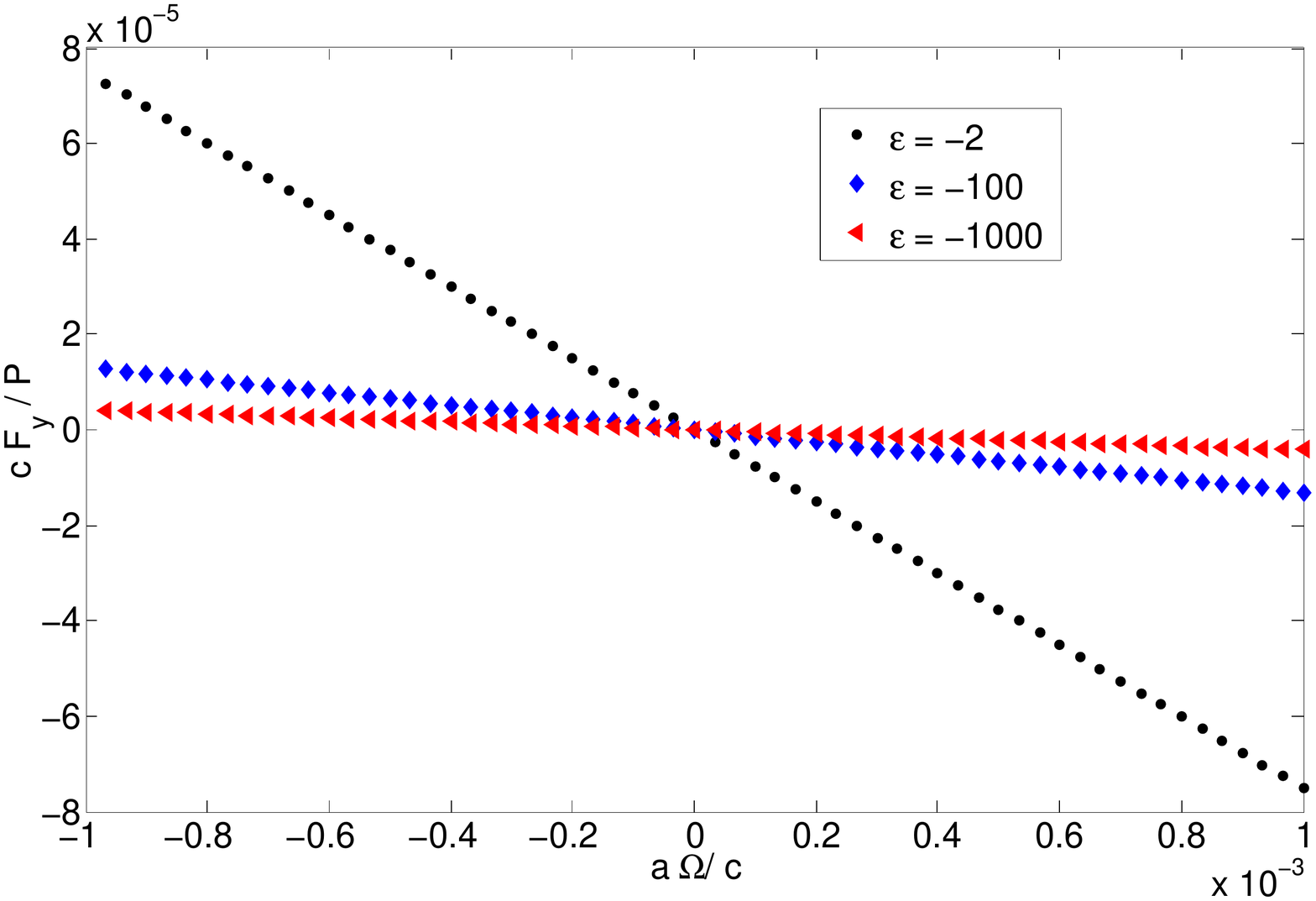}\caption{\label{fig:Metals}(Color online) Normalized Force vs. Rotational
frequency for $\lambda_{0}=0.01$ and various $\epsilon/\epsilon_{0}$
where $P=2a\left|E_{0}\right|^{2}\sqrt{\frac{\epsilon_{0}}{\mu_{0}}}$.}
\end{figure}

In the case of a perfect conductor, $\beta_{n}$ is zero (see Eq.
\ref{eq:E_t_RAW}) and therefore Eqs. \ref{eq:BC_alpha_beta} and
\ref{eq:BC_alpha_beta2} do not give an asymmetry with respect to
$n$ for $\alpha_{n}$. Consequently, in this limit there is no lateral
force. Intuitively, because a perfect conductor allows no penetration
of the electromagnetic fields, the fields cannot ``notice'' that
it is rotating or be ``dragged'' by the moving matter. However,
for imperfect metals (finite $\epsilon<0$) there is some penetration
of the radiation into the material which results in a lateral force.
Interestingly, in the case of $\epsilon<0$, and in fact whenever
$\epsilon<\epsilon_{0}$ (negative susceptibility), the force is in
the opposite direction of the force for $\epsilon>\epsilon_{0}$ dielectrics
(see Fig. \ref{fig:Metals}). The reason is an immediate consequence
of Eq. \ref{eq:K}. For $\epsilon<\epsilon_{0}$ and $\Omega>0$,
$K$ becomes negative and the phenomenology, looking at $\gamma_{n}$
in Eq.~\ref{eq:gamma_n}, become equivalent to the case of $\epsilon>\epsilon_{0}$
and $\Omega<0$. The same relationship between the sign of the force
and the sign of $\Re\epsilon-\epsilon_{0}$ holds for complex $\epsilon$
as long as $|\Im\epsilon|\ll|\Re\epsilon|$, whereas for large $|\Im\epsilon|$
we observe a similar relationship with the sign of the imaginary part.

\begin{figure}
\centering{}\includegraphics[scale=0.3]{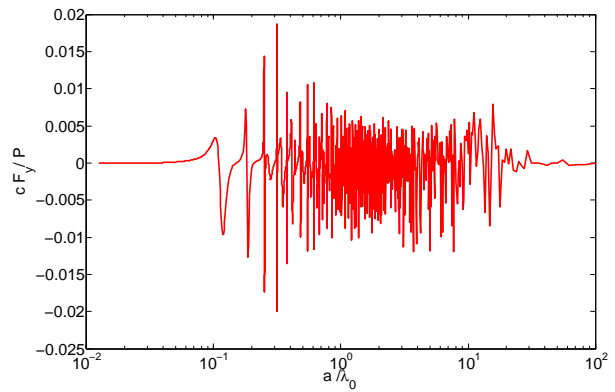}\caption{\label{fig:Force-vs.-alam}(Color online) Mie resonances: Normalized
force vs. $\lambda_{0}$ for $a\Omega/c=3.34e-4$ and $a=0.01$.}
\end{figure}

Lastly, we investigate the dependence of the normalized force on $2\pi\omega a/c=a/\lambda_{0}$,
varying the vacuum wavelength $\lambda_{0}$ (see Fig. \ref{fig:Force-vs.-alam}).
For $\lambda_{0}\ll a$ the scattering approaches a ray-optics limit,
while for $\lambda_{0}\gg a$ it is in the Rayleigh-scattering (dipole
approximation) regime~\cite{Jackson1998}. For $\lambda_{0}\sim a$,
the force spectrum becomes more interesting due to the presence of
Mie resonances~\cite{Stratton1941}.

\textit{\label{sec:Discussion-and-Future}Discussion and Future Work.--
} Given a finite amount of power, one would use a focused beam rather
than a plane-wave, and an interesting question for future work is
what beam width (and profile) maximizes the lateral force for a given
total power; we conjecture that the optimal beam width should be comparable
to the scattering cross-section.

Furthermore, recent work has shown that an appropriate beam can form
an optical \textquotedbl{}tweezers\textquotedbl{} \cite{ashkin1986}
or \textquotedbl{}tractor beam\textquotedbl{} in which the sign of
the longitudinal force on a non-spinning particle can reverse \cite{Chen2011,lee2010,Marston2006}.
Applied to a spinning particle, the ability to change the sign of
the longitudinal force implies that there should also be a zero point:
a beam for which the force of is \emph{purely lateral}.

The forces obtained here are only a fraction of the incident radiation
pressure and seem to require infeasible rotation rates, but we expect
that they can be resonantly enhanced by techniques similar to those
that have been used by other authors to enhance scattered power for
a given particle diameter. Mie resonances are already visible in Fig.~\ref{fig:Force-vs.-alam},
but much stronger resonant phenomena can be designed by using multilayer
spheres that trap light using Bragg mirrors and/or specially designed
surface plasmons, and one can even obtain ``superscattering'' by
aligning multiple resonances at the same frequency~\cite{Ruan11}.

Material dispersion will contribute an additional source of lateral
force: similar to the origin of quantum friction ~\cite{Zhao12,Manjavacas10,Pendry10},
the Doppler shift in the material dispersion should differ between
the sides of the object moving toward and away from the light source,
causing additional asymmetry in the scattered field and hence additional
lateral force.

Such enhancement mechanism, in combination with recent progress in
generating rotating particles (of graphene) at near-GHz $\Omega$
\cite{kane2010}, may permit the future experimental observation and
exploitation of optical ``Bernoulli'' forces.

SGJ was supported in part by the U.S. Army Research Office under contract
W911NF-13-D-0001.

\bibliographystyle{apsrev4-1}
\bibliography{mybib}

\newpage{}

\section{Appendix}

First let $\sigma,\mu,\epsilon$ be the conductivity, permeability
and permittivity respectively and define $\mathbf{\Lambda}=\left(\epsilon\mu-\epsilon_{0}\mu_{0}\right)\vec{v}$,
where subscript $0$ corresponds to quantities in the vacuum. The
electric and magnetic field vectors satisfy

\begin{singlespace}
\begin{eqnarray}
\nabla\times\vec{E} & = & -\frac{\partial}{\partial t}\left(\mu\vec{H-\Lambda}\times\vec{E}\right)\label{eq:EM_Tai-1}\\
\nabla\times\vec{H} & = & \sigma\left(\vec{E}+\mu\vec{v}\times\vec{H}\right)+\frac{\partial}{\partial t}\left(\sigma\vec{E+\Lambda}\times\vec{H}\right).\label{eq:EMB_Tai-1}
\end{eqnarray}

\end{singlespace}

To determine the proper expression for the transmitted wave, we use
harmonically oscillating fields to reduce Eqs. \ref{eq:EM_Tai-1}
and \ref{eq:EMB_Tai-1} to (neglecting $\mathcal{O}\left(\left(v/c\right)^{2}\right)$
terms)

\begin{singlespace}
\begin{eqnarray}
\left(\nabla+i\omega\vec{\Lambda}\right)\times\vec{E} & = & i\omega\mu\vec{H}\label{eq:HarmonicReduction_1}\\
\left(\nabla+i\omega\vec{\Lambda}-\sigma\mu\vec{v}\right)\times\vec{H} & = & \left(\sigma-i\omega\epsilon\right)\vec{E}\label{eq:HarmonicReduction_2}
\end{eqnarray}

Looking at Figure \ref{fig:CylindricalGeometry}, the incident wave
is given by

\begin{equation}
E_{i}=E_{0}\mbox{exp}\left\{ ik_{0}r\cos\phi\right\} =E_{0}\sum_{n=-\infty}^{+\infty}i^{n}J_{n}\left(k_{0}r\right)\mbox{exp}\left(in\phi\right),\label{eq:IncidentE_Tai-1}
\end{equation}
 where $k_{0}=\omega/c$ is the wave number in vacuo; $\omega$ being
the frequency in the time harmonic oscillating field $e^{i\omega t}$.
The scattering field can be written in the form

\begin{equation}
E_{s}=E_{0}\sum_{n=-\infty}^{+\infty}\alpha_{n}i^{n}H_{n}^{\left(1\right)}\left(k_{0}r\right)\mbox{exp}\left(in\phi\right).\label{eq:ScatteredE_Tai-1}
\end{equation}
 To solve for the transmitted field inside the dielectric we subject
Eqs. \ref{eq:HarmonicReduction_1} and \ref{eq:HarmonicReduction_2}
to the particular form of the velocity which is independent of $\mathbf{\hat{z}}$.
The parameter $\vec{\Lambda}$ is then a function of $r$ alone,

\begin{equation}
\vec{\Lambda=}\mu_{0}\left(\epsilon-\epsilon_{0}\right)\vec{v}=Kr\mathbf{\hat{\vec{\phi}}}\label{eq:Lambda-1}
\end{equation}
 where $K\equiv\mu_{0}\epsilon\left(1-\frac{\epsilon_{0}}{\epsilon}\right)\Omega=\frac{m\Omega}{c^{2}}$,
with $m\equiv1-\frac{\epsilon_{0}}{\epsilon}$ and $c^{2}=1/\mu_{0}\epsilon$.
Substituting Eq. \ref{eq:Lambda-1} into Eqs. \ref{eq:HarmonicReduction_1}
and \ref{eq:HarmonicReduction_2} and eliminating $\vec{H}$ we obtain
a differential equation for $E_{z}$ which is the only component of
the electric field inside the dielectric cylinder.

\begin{equation}
\frac{1}{r}\frac{\partial}{\partial r}\left(r\frac{\partial E_{z}}{\partial r}\right)+\frac{1}{r^{2}}\frac{\partial^{2}E_{z}}{\partial\phi^{2}}+2i\omega K\frac{\partial E_{z}}{\partial\phi}+k^{2}E_{z}+\mathcal{O}\left(\left(\frac{v}{c}\right)^{2}\right)=0,\label{eq:DiffEq_Ez-1}
\end{equation}
 where $k^{2}=\left(\omega/c\right)^{2}$. To solve let us seek separable
solutions for $E_{z}=F\left(r\right)e^{in\phi}$ and below we drop
$\mathcal{O}\left(\left(\frac{v}{c}\right)^{2}\right)$ by understanding
that the results are accurate to first order. The function $F\left(r\right)$
then satisfies

\[
\frac{1}{r}\frac{\partial}{\partial r}\left(r\frac{\partial F}{\partial r}\right)-\left(\frac{n^{2}}{r^{2}}+2n\omega K-k^{2}\right)F=0.
\]

If we introduce $\gamma_{n}^{2}=k^{2}-2n\omega K=k^{2}\left(1-\frac{2nm\Omega}{\omega}\right),$
then the proper set of radial functions to describe the field inside
the rotating cylinder is

\[
F\left(r\right)=J_{n}\left(\gamma_{n}r\right),\qquad n=0,\pm1,\pm2,\cdots.
\]
 The complete expression for the transmitted field can be written
in the form

\begin{equation}
E_{t}=E_{0}\sum_{n=-\infty}^{+\infty}\beta_{n}i^{n}J_{n}\left(\gamma_{n}r\right)\mbox{exp}\left(in\phi\right).\label{eq:E_t_RAW-1}
\end{equation}
 By matching $E_{z}$ as defined by Eqs. \ref{eq:IncidentE_Tai-1},
\ref{eq:ScatteredE_Tai-1} and \ref{eq:E_t_RAW-1}, and the $\phi-$component
of the magnetic field at the boundary $r=a$ one obtains the following
two simultaneous equations:

\begin{eqnarray}
J_{n}\left(k_{0}a\right)+\alpha_{n}H_{n}^{\left(1\right)}\left(k_{0}a\right) & = & \beta_{n}J_{n}\left(\gamma_{n}a\right)\label{eq:BC_alpha_beta-1}\\
k_{0}\left[J_{n}'\left(k_{0}a\right)+\alpha_{n}H_{n}^{\left(1\right)}\left(k_{0}a\right)'\right] & = & \beta_{n}\gamma_{n}J_{n}'\left(\gamma_{n}a\right),
\end{eqnarray}
 where the prime on $J$ and $H^{\left(1\right)}$ denotes derivative
with respect to the entire argument of these functions. The solutions
for $\alpha_{n}$ and $\beta_{n}$ are

\begin{eqnarray}
\alpha_{n} & = & -\frac{J_{n}\left(\rho_{0}\right)J'_{n}\left(\rho_{n}\right)-\frac{k_{0}}{\gamma_{n}}J_{n}\left(\rho_{n}\right)J'_{n}\left(\rho_{0}\right)}{H_{n}^{\left(1\right)}\left(\rho_{0}\right)J'_{n}\left(\rho_{n}\right)-\frac{k_{0}}{\gamma_{n}}J_{n}\left(\rho_{n}\right)H_{n}^{\left(1\right)}\mbox{}'\left(\rho_{0}\right)}\label{eq:alpha_n-1}\\
\beta_{n} & = & -\frac{k_{0}}{\gamma_{n}}\frac{\left[J_{n}\left(\rho_{0}\right)H_{n}^{\left(1\right)}\mbox{}'\left(\rho_{0}\right)-J_{n}'\left(\rho_{0}\right)H_{n}^{\left(1\right)}\left(\rho_{0}\right)\right]}{H_{n}^{\left(1\right)}\left(\rho_{0}\right)J'_{n}\left(\rho_{n}\right)-\frac{k_{0}}{\gamma_{n}}J_{n}\left(\rho_{n}\right)H_{n}^{\left(1\right)}\mbox{}'\left(\rho_{0}\right)},\label{eq:beta_n-1}
\end{eqnarray}
 where $\rho_{0}=k_{0}a$ and $\rho_{n}=\gamma_{n}a$. Using identities
$J'_{n}\left(x\right)=\frac{1}{2}\left[J_{n-1}\left(x\right)-J_{n+1}\left(x\right)\right]$,
$J'_{0}\left(x\right)=-J_{1}\left(x\right)$ and $H_{n}^{\left(1\right)}\mbox{}'\left(x\right)=\frac{1}{2}\left[H_{n-1}^{\left(1\right)}\left(x\right)-H_{n+1}^{\left(1\right)}\left(x\right)\right]$,
$H_{0}^{\left(1\right)}\mbox{}'\left(x\right)=-H_{1}^{\left(1\right)}\left(x\right)$,
$\beta_{n}$ can be eliminated to give 
\end{singlespace}

\begin{widetext}

\begin{equation}
\alpha_{n}=-\frac{J_{n}\left(\rho_{0}\right)\left[J_{n-1}\left(\rho_{n}\right)-J_{n+1}\left(\rho_{n}\right)\right]-\frac{k_{0}}{\gamma_{n}}J_{n}\left(\rho_{n}\right)\left[J_{n-1}\left(\rho_{0}\right)-J_{n+1}\left(\rho_{0}\right)\right]}{H_{n}^{\left(1\right)}\left(\rho_{0}\right)\left[J_{n-1}\left(\rho_{n}\right)-J_{n+1}\left(\rho_{n}\right)\right]-\frac{k_{0}}{\gamma_{n}}J_{n}\left(\rho_{n}\right)\left[H_{n-1}^{\left(1\right)}\left(\rho_{0}\right)-H_{n+1}^{\left(1\right)}\left(\rho_{0}\right)\right]}.\label{eq:alphan_simplified-1}
\end{equation}

\end{widetext}

\begin{singlespace}
The numerical value of $\alpha_{n}\ne\alpha_{-n}$ for $\Omega\ne0$
because in this case $\rho_{n}\ne\rho_{-n}$ and $\gamma_{n}\ne\gamma_{-n}$
hence the scattering field has an asymmetrical part with respect to
the direction of incidence, $\phi=0$. When $\Omega=0\implies\alpha_{n}=\alpha_{-n}$
and Eq. \ref{eq:ScatteredE_Tai-1} reduces to the well known results. 
\end{singlespace}

The total field therefore is $\mathbf{E}=E_{z}\hat{\mathbf{z}}=\left(E_{i}+E_{s}\right)\hat{\mathbf{z}}$.
Below we suppress $k_{0}r$ as the argument of the Bessel functions
unless stated otherwise.

\begin{equation}
E_{z}=E_{0}\sum_{n=-\infty}^{+\infty}i^{n}\left\{ J_{n}+\alpha_{n}H_{n}^{\left(1\right)}\right\} e^{in\phi}\label{eq:Ez-1-2}
\end{equation}

Further in free space we have $\vec{H}=\frac{1}{i\omega\mu_{0}}\nabla\times\vec{E}$
which gives

\begin{eqnarray}
\vec{H} & = & \frac{1}{i\omega\mu_{0}}\nabla\times E\hat{\vec{z}}=\frac{1}{i\omega\mu_{0}}\left(\frac{1}{r}\frac{\partial E_{z}}{\partial\phi}\hat{\vec{r}}-\frac{\partial E_{z}}{\partial r}\hat{\vec{\phi}}\right)\label{eq:H-1}\\
 & = & \frac{1}{i\omega\mu_{0}}\left(\frac{1}{r}\frac{\partial E_{z}}{\partial\phi}\cos\phi+\frac{\partial E_{z}}{\partial r}\sin\phi\right)\hat{\vec{x}}\nonumber \\
 & + & \frac{1}{i\omega\mu_{0}}\left(\frac{1}{r}\frac{\partial E_{z}}{\partial\phi}\sin\phi-\frac{\partial E_{z}}{\partial r}\cos\phi\right)\hat{\vec{y}}.\nonumber 
\end{eqnarray}

\begin{singlespace}
Let $\vec{\hat{n_{0}}=}\cos\psi\vec{\hat{x}}+\sin\psi\vec{\hat{y}}$
be any unit vector, then Eq. \ref{eq:StressTensorSI}, evaluated at
the radius $r_{0}$, reads

\begin{eqnarray}
F_{\vec{\hat{n_{0}}}} & = & r_{0}\oint d\phi\left\{ -\frac{\epsilon_{0}\left|E\right|^{2}+\mu_{0}\left|H\right|^{2}}{4}\cos\left(\psi-\phi\right)\right.\label{eq:ForceGeneralEz}\\
 & + & \frac{\mu_{0}}{2}\mbox{Re}\left(H_{x}^{*}H_{y}\right)\sin\left(\psi+\phi\right)\nonumber \\
 & + & \left.\frac{\mu_{0}\left|H_{x}\right|^{2}}{2}\cos\phi\cos\psi+\frac{\mu_{0}\left|H_{y}\right|^{2}}{2}\sin\phi\sin\psi\right\} .\nonumber 
\end{eqnarray}

In particular we are interested in $\psi=\frac{\pi}{2}$ to calculate
the transverse force

\begin{eqnarray}
F_{\vec{\hat{y}}} & = & r_{0}\oint d\phi\left\{ -\frac{\epsilon_{0}\left|E\right|^{2}+\mu_{0}\left|H\right|^{2}}{4}\sin\phi\right.\label{eq:ForceY-1}\\
 & + & \left.\frac{\mu_{0}}{2}\mbox{Re}\left(H_{x}^{*}H_{y}\right)\cos\phi+\frac{\mu_{0}\left|H_{y}\right|^{2}}{2}\sin\phi\right\} .
\end{eqnarray}
 where $E=\left(E_{i}+E_{s}\right)$ is given by Eqs. \ref{eq:IncidentE_Tai}
and \ref{eq:ScatteredE_Tai}. 
\end{singlespace}

\section{Evaluating the Integral}

Here we evaluate $\left|E\right|^{2}$, $\left|H\right|^{2}$, $\left|H_{y}^{2}\right|$
and $\mbox{Re}\left(H_{x}^{*}H_{y}\right)$ as they are useful for
calculating the force below (Eqs. \ref{eq:ForceGeneralEz} and \ref{eq:ForceY}).
The key equations are

\begin{eqnarray*}
E_{z} & = & E_{0}\sum_{n=-\infty}^{+\infty}i^{n}\left\{ J_{n}+\alpha_{n}H_{n}^{\left(1\right)}\right\} e^{in\phi}\\
\vec{H} & \equiv & H_{x}\hat{\vec{x}}+H_{y}\hat{\vec{y}}\\
 & = & \frac{1}{i\omega\mu_{0}}\left(\frac{1}{r}\frac{\partial E_{z}}{\partial\phi}\cos\phi+\frac{\partial E_{z}}{\partial r}\sin\phi\right)\hat{\vec{x}}\\
 & + & \frac{1}{i\omega\mu_{0}}\left(\frac{1}{r}\frac{\partial E_{z}}{\partial\phi}\sin\phi-\frac{\partial E_{z}}{\partial r}\cos\phi\right)\hat{\vec{y}}.
\end{eqnarray*}
 where

\begin{widetext}

\begin{eqnarray*}
H_{x} & = & \frac{E_{0}}{i\omega\mu_{0}}\sum_{n=-\infty}^{+\infty}e^{in\phi}i^{n}\left\{ \frac{in}{r}\left[J_{n}+\alpha_{n}H_{n}^{\left(1\right)}\right]\cos\phi+\frac{k_{0}}{2}\left[J_{n-1}-J_{n+1}+\alpha_{n}\left(H_{n-1}^{\left(1\right)}-H_{n+1}^{\left(1\right)}\right)\right]\sin\phi\right\} \\
H_{y} & = & \frac{E_{0}}{i\omega\mu_{0}}\sum_{n=-\infty}^{+\infty}e^{in\phi}i^{n}\left\{ \frac{in}{r}\left[J_{n}+\alpha_{n}H_{n}^{\left(1\right)}\right]\sin\phi-\frac{k_{0}}{2}\left[J_{n-1}-J_{n+1}+\alpha_{n}\left(H_{n-1}^{\left(1\right)}-H_{n+1}^{\left(1\right)}\right)\right]\cos\phi\right\} 
\end{eqnarray*}
 \end{widetext}as well as the orthogonality relations

\begin{widetext}

\begin{eqnarray*}
\oint d\phi e^{i\left(n-m\right)\phi}\sin\phi & = & \left(-i\pi\right)\left(\delta_{n,m-1}-\delta_{n,m+1}\right)\\
\oint d\phi e^{i\left(n-m\right)\phi}\sin^{3}\phi & = & \left(\frac{i\pi}{4}\right)\left(\delta_{n,m-3}-\delta_{n,m+3}-3\delta_{n,m-1}+3\delta_{n,m+1}\right)\\
\oint d\phi e^{i\left(n-m\right)\phi}\sin^{2}\phi\cos\phi & = & \left(\frac{-\pi}{4}\right)\left(\delta_{n,m-3}+\delta_{n,m+3}-\delta_{n,m-1}-\delta_{n,m+1}\right)\\
\oint d\phi e^{i\left(n-m\right)\phi}\sin\phi\cos^{2}\phi & = & \left(\frac{-i\pi}{4}\right)\left(\delta_{n,m-3}-\delta_{n,m+3}+\delta_{n,m-1}-\delta_{n,m+1}\right)\\
\oint d\phi e^{i\left(n-m\right)\phi}\cos^{3}\phi & = & \left(\frac{\pi}{4}\right)\left(\delta_{n,m-3}+\delta_{n,m+3}+3\delta_{n,m-1}+3\delta_{n,m+1}\right).
\end{eqnarray*}

\end{widetext}

We proceed

\begin{widetext}

\begin{eqnarray*}
\oint d\phi\left|E_{z}\right|^{2}\sin\phi & = & E_{0}^{2}\sum_{n,m=-\infty}^{+\infty}\oint d\phi e^{i\left(n-m\right)\phi}\sin\phi\left(-1\right)^{m}i^{n+m}\left\{ J_{n}+\alpha_{n}H_{n}^{\left(1\right)}\right\} \left\{ J_{m}+\alpha_{m}^{*}H_{m}^{\left(2\right)}\right\} \\
 & = & E_{0}^{2}\sum_{n,m=-\infty}^{+\infty}\left(-1\right)^{m}i^{n+m}\left\{ J_{n}+\alpha_{n}H_{n}^{\left(1\right)}\right\} \left\{ J_{m}+\alpha_{m}^{*}H_{m}^{\left(2\right)}\right\} \left(-i\pi\right)\left\{ \delta_{n,m-1}-\delta_{n,m+1}\right\} 
\end{eqnarray*}

\end{widetext}Further $\oint d\phi\left|H_{x}\right|^{2}\sin\phi=\oint d\phi\left(H_{x}^{*}H_{x}\right)\sin\phi$,
similarly for $\oint d\phi\left|H_{y}\right|^{2}\sin\phi$

\begin{widetext}

\begin{eqnarray*}
\oint d\phi\left|H_{x}\right|^{2}\sin\phi & = & \left(\frac{E_{0}}{\omega\mu_{0}}\right)^{2}\sum_{n,m=-\infty}^{+\infty}\left(-1\right)^{m}i^{n+m}\oint d\phi\sin\phi e^{i\left(n-m\right)\phi}\\
 &  & \left\{ \left(\frac{-im}{r}\left[J_{m}+\alpha_{m}^{*}H_{m}^{\left(2\right)}\right]\cos\phi+\frac{k_{0}}{2}\left[J_{m-1}-J_{m+1}+\alpha_{m}^{*}\left(H_{m-1}^{\left(2\right)}-H_{m+1}^{\left(2\right)}\right)\right]\sin\phi\right)\right.\\
 & \times & \left.\left(\frac{in}{r}\left[J_{n}+\alpha_{n}H_{n}^{\left(1\right)}\right]\cos\phi+\frac{k_{0}}{2}\left[J_{n-1}-J_{n+1}+\alpha_{n}\left(H_{n-1}^{\left(1\right)}-H_{n+1}^{\left(1\right)}\right)\right]\sin\phi\right)\right\} 
\end{eqnarray*}

\end{widetext}which using the orthogonality relations yields

\begin{widetext}

\begin{eqnarray*}
\oint d\phi\left|H_{x}\right|^{2}\sin\phi & = & \left(\frac{E_{0}}{\omega\mu_{0}}\right)^{2}\sum_{n,m=-\infty}^{+\infty}\left(-1\right)^{m}i^{n+m}\\
 &  & \frac{nm}{r^{2}}\left(J_{m}+\alpha_{m}^{*}H_{m}^{\left(2\right)}\right)\left(J_{n}+\alpha_{n}H_{n}^{\left(1\right)}\right)\left(\frac{-i\pi}{4}\right)\left(\delta_{n,m-3}-\delta_{n,m+3}+\delta_{n,m-1}-\delta_{n,m+1}\right)\\
 & + & \frac{k_{0}^{2}}{4}\left[J_{m-1}-J_{m+1}+\alpha_{m}^{*}\left(H_{m-1}^{\left(2\right)}-H_{m+1}^{\left(2\right)}\right)\right]\left[J_{n-1}-J_{n+1}+\alpha_{n}\left(H_{n-1}^{\left(1\right)}-H_{n+1}^{\left(1\right)}\right)\right]\\
 & \times & \left(\frac{i\pi}{4}\right)\left(\delta_{n,m-3}-\delta_{n,m+3}-3\delta_{n,m-1}+3\delta_{n,m+1}\right)\\
 & + & \frac{ik_{0}}{2r}\left[-m\left(J_{m}+\alpha_{m}^{*}H_{m}^{\left(2\right)}\right)\left(J_{n-1}-J_{n+1}+\alpha_{n}\left(H_{n-1}^{\left(1\right)}-H_{n+1}^{\left(1\right)}\right)\right)\right.\\
 &  & \;\left.+n\left(J_{n}+\alpha_{n}H_{n}^{\left(1\right)}\right)\left(J_{m-1}-J_{m+1}+\alpha_{m}^{*}\left(H_{m-1}^{\left(2\right)}-H_{m+1}^{\left(2\right)}\right)\right)\right]\\
 & \times & \left(\frac{-\pi}{4}\right)\left(\delta_{n,m-3}+\delta_{n,m+3}-\delta_{n,m-1}-\delta_{n,m+1}\right)
\end{eqnarray*}

\end{widetext}

Similarly \begin{widetext}

\begin{eqnarray*}
\oint d\phi\left|H_{y}\right|^{2}\sin\phi & = & \left(\frac{E_{0}}{\omega\mu_{0}}\right)^{2}\sum_{n,m=-\infty}^{+\infty}\left(-1\right)^{m}i^{n+m}\oint d\phi\sin\phi e^{i\left(n-m\right)\phi}\\
 &  & \left\{ \left(\frac{-im}{r}\left[J_{m}+\alpha_{m}^{*}H_{m}^{\left(2\right)}\right]\sin\phi-\frac{k_{0}}{2}\left[J_{m-1}-J_{m+1}+\alpha_{m}^{*}\left(H_{m-1}^{\left(2\right)}-H_{m+1}^{\left(2\right)}\right)\right]\cos\phi\right)\right.\\
 & \times & \left.\left(\frac{in}{r}\left[J_{n}+\alpha_{n}H_{n}^{\left(1\right)}\right]\sin\phi-\frac{k_{0}}{2}\left[J_{n-1}-J_{n+1}+\alpha_{n}\left(H_{n-1}^{\left(1\right)}-H_{n+1}^{\left(1\right)}\right)\right]\cos\phi\right)\right\} 
\end{eqnarray*}

\end{widetext}which using the orthogonality relations yields

\begin{widetext}

\begin{eqnarray*}
\oint d\phi\left|H_{y}\right|^{2}\sin\phi & = & \left(\frac{E_{0}}{\omega\mu_{0}}\right)^{2}\sum_{n,m=-\infty}^{+\infty}\left(-1\right)^{m}i^{n+m}\\
 &  & \frac{mn}{r^{2}}\left(J_{n}+\alpha_{n}H_{n}^{\left(1\right)}\right)\left(J_{m}+\alpha_{m}^{*}H_{m}^{\left(2\right)}\right)\left(\frac{i\pi}{4}\right)\left(\delta_{n,m-3}-\delta_{n,m+3}-3\delta_{n,m-1}+3\delta_{n,m+1}\right)\\
 & + & \frac{k_{0}^{2}}{4}\left(J_{m-1}-J_{m+1}+\alpha_{m}^{*}\left(H_{m-1}^{\left(2\right)}-H_{m+1}^{\left(2\right)}\right)\right)\left(J_{n-1}-J_{n+1}+\alpha_{n}\left(H_{n-1}^{\left(1\right)}-H_{n+1}^{\left(1\right)}\right)\right)\\
 & \times & \left(\frac{-i\pi}{4}\right)\left(\delta_{n,m-3}-\delta_{n,m+3}+\delta_{n,m-1}-\delta_{n,m+1}\right)\\
 & + & \frac{ik_{0}}{2r}\left[m\left(J_{m}+\alpha_{m}^{*}H_{m}^{\left(2\right)}\right)\left(J_{n-1}-J_{n+1}+\alpha_{n}\left(H_{n-1}^{\left(1\right)}-H_{n+1}^{\left(1\right)}\right)\right)\right.\\
 &  & \;\left.-n\left(J_{n}+\alpha_{n}H_{n}^{\left(1\right)}\right)\left(J_{m-1}-J_{m+1}+\alpha_{m}^{*}\left(H_{m-1}^{\left(2\right)}-H_{m+1}^{\left(2\right)}\right)\right)\right]\\
 & \times & \left(\frac{-\pi}{4}\right)\left(\delta_{n,m-3}+\delta_{n,m+3}-\delta_{n,m-1}-\delta_{n,m+1}\right).
\end{eqnarray*}
 \end{widetext}

Lastly we need $\oint d\phi\mbox{Re}\left(H_{x}^{*}H_{y}\right)\cos\phi=\frac{1}{2}\oint d\phi H_{x}^{*}H_{y}\cos\phi+c.c.$

\begin{widetext}

\begin{eqnarray*}
\oint d\phi H_{x}^{*}H_{y}\cos\phi & = & \left(\frac{E_{0}}{\omega\mu_{0}}\right)^{2}\sum_{n,m=-\infty}^{+\infty}\left(-1\right)^{m}i^{n+m}\oint d\phi\cos\phi e^{i\left(n-m\right)\phi}\\
 &  & \left\{ \left(\frac{-im}{r}\left[J_{m}+\alpha_{m}^{*}H_{m}^{\left(2\right)}\right]\cos\phi+\frac{k_{0}}{2}\left[J_{m-1}-J_{m+1}+\alpha_{m}^{*}\left(H_{m-1}^{\left(2\right)}-H_{m+1}^{\left(2\right)}\right)\right]\sin\phi\right)\right.\\
 & \times & \left.\left(\frac{in}{r}\left[J_{n}+\alpha_{n}H_{n}^{\left(1\right)}\right]\sin\phi-\frac{k_{0}}{2}\left[J_{n-1}-J_{n+1}+\alpha_{n}\left(H_{n-1}^{\left(1\right)}-H_{n+1}^{\left(1\right)}\right)\right]\cos\phi\right)\right\} 
\end{eqnarray*}

\end{widetext}which using the orthogonality relations yields

\begin{widetext}

\begin{eqnarray*}
\oint d\phi H_{x}^{*}H_{y}\cos\phi & = & \left(\frac{E_{0}}{\omega\mu_{0}}\right)^{2}\sum_{n,m=-\infty}^{+\infty}\left(-1\right)^{m}i^{n+m}\\
 &  & \left[\frac{mn}{r^{2}}\left(J_{m}+\alpha_{m}^{*}H_{m}^{\left(2\right)}\right)\left(J_{n}+\alpha_{n}H_{n}^{\left(1\right)}\right)-\frac{k_{0}^{2}}{4}\left(J_{m-1}-J_{m+1}+\alpha_{m}^{*}\left(H_{m-1}^{\left(2\right)}-H_{m+1}^{\left(2\right)}\right)\right)\right.\\
 &  & \left.\times\left(J_{n-1}-J_{n+1}+\alpha_{n}\left(H_{n-1}^{\left(1\right)}-H_{n+1}^{\left(1\right)}\right)\right)\left(-\frac{i\pi}{4}\right)\right]\left(\delta_{n,m-3}-\delta_{n,m+3}+\delta_{n,m-1}-\delta_{n,m+1}\right)\\
 & + & \frac{imk_{0}}{2r}\left(J_{m}+\alpha_{m}^{*}H_{m}^{\left(2\right)}\right)\left(J_{n-1}-J_{n+1}+\alpha_{n}\left(H_{n-1}^{\left(1\right)}-H_{n+1}^{\left(1\right)}\right)\right)\\
 &  & \times\left(\frac{\pi}{4}\right)\left(\delta_{n,m-3}+\delta_{n,m+3}+3\delta_{n,m-1}+3\delta_{n,m+1}\right)\\
 & + & \frac{ink_{0}}{2r}\left(J_{n}+\alpha_{n}H_{n}^{\left(1\right)}\right)\left(J_{m-1}-J_{m+1}+\alpha_{m}^{*}\left(H_{m-1}^{\left(2\right)}-H_{m+1}^{\left(2\right)}\right)\right)\\
 &  & \times\left(\frac{-\pi}{4}\right)\left(\delta_{n,m-3}+\delta_{n,m+3}-\delta_{n,m-1}-\delta_{n,m+1}\right).
\end{eqnarray*}

\end{widetext}

All of the above integrals were checked against numerics before calculating
the cumulative effect that appears in the force Eq. \ref{eq:ForceY}.
The total force was also checked against numerical experiments. In
all cases agreements were found with errors of order $\mathcal{O}\left(10^{-26}\right)$. 
\end{document}